\documentstyle[aps,prd,preprint,tighten,floats,epsf]{revtex}

\def \GeV {{\rm GeV}}

\def \mb {{\rm mb}}

\begin{document}
%

\title{Jet Production Cross Section with Double Pomeron Exchange}

\author{Arjun Berera}

\address{
   Department of Physics, 
   Pennsylvania State University,
   University Park, PA 16802, U.S.A.
   August 1995
}

\maketitle

\begin{abstract}
I will discuss hadron-hadron collisions where the final state is
kinematically of the kind associated with double-pomeron-exchange (DPE)
and has large transverse momentum jets.
I will show that in addition to the
conventional factorized (FDPE) contribution, there is a
non-factorized (NDPE) contribution which has no pomeron beam jet.
Results of calculations from a simple model of DPE for
two-jet total and differential cross
sections at Tevatron energy scales are presented. The NDPE
contribution is dominant.
\end{abstract}

\medskip

1995 Blois Conference

\clearpage
\setlength{\baselineskip}{3.2ex}

An interesting class of diffractive hadron-hadron collisions, is where
{\em both} hadrons survive unscathed, but leave a remnant system in
the central region of final-state rapidity.  Such events are called
double-pomeron exchange (DPE) events \cite {DPEthy,DPEexp}.  In
effect, one can try to consider such processes as pomeron-pomeron
collisions.  In this talk I will discuss the properties of
jet production in DPE events, and in particular the breakdown of
hard-scattering factorization.
Jet production by DPE has been reported \cite{DPE.jet.exp} in the UA1
detector.

I will discuss how DPE
at the level of lowest-order Feynman graphs in a crude model
provides a striking mechanism for the breakdown of factorization.
The model is in
effect a version of the Low-Nussinov-Gunion-Soper
model \cite{LowNussinov}, and the same model
was used by Berera and Soper
\cite{bersop} to understand properties of the pomeron's structure
function.  I will present calculations done with J. C. Collins
on the cross section for jet production in
DPE \cite{bercol}.  The important free parameter in our model is an overall
normalization which can be determined from elastic scattering.
We call the model non-factorizing double-pomeron-exchange (NDPE), to
contrast it with the Ingelman-Schlein model applied to DPE
\cite{IS}, which we
call factorized double-pomeron-exchange (FDPE). The theoretical
issues of nonfactorization are discussed in [8].

The dramatic feature of our model is that the pomerons have no beam
jets; the final state is then exceptionally clean, because it consists
of the two isolated, diffracted hadrons, the high transverse-momentum
jets, and nothing else.
Not only are such processes theoretically interesting in their own
right, but they have advantages for certain kinds of new particle
searches, provided the cross section is high enough, because a lot of
the background event is no longer present.  Production processes
of heavy flavors and Higgs are two such examples.
But studies of hard scattering in DPE should surely
start with the simplest processes, jet production.

The process I am interested in is
\begin{equation}A+B\rightarrow A'+B'+\mbox{\rm 2 jets}.
\label{DPE.jets}
\end{equation}
The hadrons $A$ and $B$ lose tiny fractions $x_a$ and $x_b$ of their
respective longitudinal momenta,
and they acquire transverse momenta ${\bf Q}_1$
and ${\bf Q}_2$.  (This defines a diffractive regime, and in Regge
theory would lead to an expectation of the dominance of
double pomeron exchange (DPE) --- 
Fig. 1)
The jets carry large momenta of magnitude
$E_T$ in the plane perpendicular to the collision axis
with azimuthal angle $\phi$.  (This defines a hard-scattering
regime.)
The small transfer of
longitudinal momentum to the hard process implies large rapidity
gaps between the jets and the two outgoing hadrons.  In what follows,
we will generically denote the hadronic scale
by $m$.
The relevant kinematic region of interest is
where $m$, ${\bf Q}_1$, and ${\bf Q}_2$ are of a typical
hadronic scale (less than about
1 GeV), while $E_T$ is much greater than this scale, but
much less than $\sqrt s$.  We want to take the limit
$x_a,x_b\rightarrow 0$, to give a
Regge-style limit, and $E_T\to\infty$, to give a hard
scattering limit.  We hold $\kappa$, ${\bf Q}_1$, and ${\bf Q}_2$ fixed, to
correspond to a fixed angle for the hard scattering and fixed
transverse momenta for the outgoing hadrons.
In this limit, $s$ necessarily goes to infinity, since it is
proportional to $E_T^2/x_ax_b$.

The process is modeled by the lowest order Feynman
graphs that are appropriate.  Our model for NDPE at lowest order
is shown in Fig. 1. The pomerons are replaced
by two-gluon exchanges, while instead of true bound
states, we model the hadrons by elementary color-singlet
scalar particles that we will call ``mesons'' and that are
coupled to scalar quarks by a $\phi^3$ coupling.  The remaining graphs
involve all possible attachments of the gluons to the quark loops.
We normalize the coupling of our mesons so that we
reproduce the measured value for cross section for small
angle elastic scattering of protons, when we use
two-gluon exchange for elastic scattering.  The details of
the calculation can be found in [4].
To compute the amplitude in Fig. 1,
one first performs contour integrals for the longitudinal
components $k^{-}$ and $k^{+}$ of gluon loop momentum, while
making the appropriate approximations to get the leading
power behavior in the limit we are considering.
The expressions we obtain for the scattering amplitude is,
\begin{eqnarray}
{\cal M}&=&-\frac {(-i)}{x_ax_b}\int\frac {d^2{\bf k}}{(2\pi )^2}\,
\hat {g}_A({\bf k},-{\bf Q}_1)\,\hat {g}_B({\bf k},{\bf Q}_2)\,\epsilon_
i({\bf k}{\bf -}{\bf Q}_{{\bf 1}})\,\epsilon_j(-{\bf k}{\bf -}{\bf Q}_{
{\bf 2}})\,{\cal A}(i,j;f)\nonumber\\
&\equiv&\frac {-i B_{ij}({\bf Q}_1,{\bf Q}_2)\,{\cal A}(i,j;f)}{x_ax_
b} ,\nonumber\\
\label{mdp}\end{eqnarray}
where ${\cal A}(i,j;f)$ is the scattering amplitude for two incoming
gluons with transverse polarization $i$ and $j$ to go
to the final state $f$, and
the ``polarization" vectors are defined as
\begin{equation}\epsilon_i({\bf k})=\frac {{\bf k}_i}{\sqrt {{\bf k}^
2}}.\label{pol}\end{equation}
This scaled amplitude has the property of being
independent of $x_a$, $x_b$, and $E_T$ in the combined Regge and
hard scattering limit that we are considering.

The retention of only the transverse components of the gluon
polarizations in ${\cal A} (\mu, \nu; f)$ is a result of
a Ward identity manipulation.  By consideration of momentum
flow, ${\cal A}(\mu,\nu;f)$ is dominated by the components in the 
directions of the initial hadrons A and B.  In light-cone coordinates,
let's say A is along the plus and B the minus directions.  This
means ${\cal A}(-,+;f)$ is the only surviving term.
However we can reexpress this term
by the QCD Ward identities and obtain:
\begin{equation}{\cal A}(-,+;f)={\cal A}(-,+;f)-\frac {p_{1'\mu}{\cal A}
(\mu ,+;f)}{p_{1'}^{+}}-\frac {{\cal A}(-,\nu ;f)p_{2'\nu}}{p_{2'}^{
-}}+\frac {p_{1'\mu}{\cal A}(\mu ,\nu ;f)p_{2'\nu}}{p_{1'}^{+}p_{
2'}^{-}}.\end{equation}
After dropping terms that are non-leading by a power,
we obtain
\begin{equation}{\cal A}(-,+;f)=\frac {p_{1'i}{\cal A}(i,j;f)p_{2'
j}}{p_{1'}^{+}p_{2'}^{-}},\label{afterWI}\end{equation}
where the Latin indices $i$ and $j$ refer to the transverse
components of vectors.
The numerators of the polarization vectors in eq. (\ref{mdp}) as
defined in eq. (\ref{pol}) arise from $p_{1'i}$ and
$p_{2'j}$ in the above manipulation.

For the cross section
in terms of jet rapidity variables we have,
\begin{equation}\frac {d\sigma}{d^2{\bf Q}_1d^2{\bf Q}_2dE^2_Td\phi_
jdy_{-}dy_{+}}=\frac {|\overline {{\cal M}}|^2\kappa^2}{2^{17}\pi^
8E_T^4}, \label{dsig2} \end{equation}
where
\begin{equation}
\kappa \equiv \frac{1}{\cosh^2 \frac{y_-}{2}}
\end{equation}
and ${\overline {{\cal M}}} \equiv x_A x_b {{\cal M}}$.
The squared amplitude can be expressed in terms of
the hadronic and hard amplitudes as follows:
\begin{equation}|\overline {{\cal M}}|^2=B^{*}_{ij}H_{ijkl}B_{kl},
\label{md2}\end{equation}
where $B_{ij}$ is given in Eq.~(\ref {mdp}) and
\begin{equation}H_{ijkl}(\kappa )=\sum_f{\cal A}^{*}(i,j;f)
{\cal A}(k,l;f).\label{H}\end{equation}
In the hard amplitude ${\cal A}(i,j;f)$, $i$ and $j$ are gluon polarization
indices while $f$ generically represents any final parton pair state.
The sum $\sum_f$ is over all spin, flavor and color states for
final-state partons of given momenta.

{}From Eqs.~(\ref{dsig2})--(\ref {md2})
at fixed $\kappa$, we see that the differential cross
section is leading twist and has the
appropriate large $s$ behavior to approximate an amplitude
with pomeron exchange.  That is, when $E_T$ gets large
and $x_a$ and $x_b$ get small, the
amplitude $\overline {{\cal M}}$ is constant,
and the cross section goes like $1/E_T^4$.

The parameters of our model are:  the mass $m$ of the
scalar quark, the mass $M$ of the mesons, the
quark-meson coupling $G$, and the gauge coupling $g$.  We
suppose that the most important unknown is the
normalization of our model for the non-perturbative
physics.  So we set all the masses to a typical hadronic
scale:  $m^2=M^2=0.1~\GeV^2$.  As for the couplings,
we have a factor $g^2$ in the hard scattering amplitude ${\cal A}$
and a factor $g^4G^4$ in the hadronic part.  The $g^2$ in
the hard amplitude should be given by the usual running
coupling at
a scale of order $E_T$. To
determine a suitable numerical value for $g^4G^4$, we apply
our model to elastic scattering.  This model is
effectively the Low-Nussinov model \cite{LowNussinov}.

We write the elastic-scattering amplitude for our model
in the Regge-like form
\begin{equation}{\cal M}_{\rm el}(t)=s\beta^2(t).\end{equation}
Here, the pomeron-hadron coupling is
given by a transverse integral,
\begin{eqnarray}
\beta^2(t)&=&-\frac{1}{4}g^4\int\frac {d^2{\bf k}}{(2\pi )^2}\hat {g}_A({\bf k}
,{\bf Q})\hat {g}_B({\bf k},{\bf Q})%
\label{bet2},\end{eqnarray}
where ${\bf Q}$ is the momentum transfer, so that
$t\equiv -{\bf Q}^2$, and $\hat {g}({\bf k},{\bf Q})$ is the same as
in eq.~(\ref {mdp}).

For forward scattering of $A$ and $B$, so when
$Q_1=Q_2=0$, there is a particularly simple form for the squared
amplitude.  It is possible to show that
\begin{equation}|\overline {{\cal M}}(0,0)|^2=64\pi\left(\frac {d\sigma
(0)}{dt}\right)_{\rm el}
\delta_{ij}\delta_{kl}H_{ijkl},\label{forcx}\end{equation}
which relates the DPE-to-jets amplitude to an elastic
cross section and a hard scattering amplitude
The elastic cross section is observable, and hard
scattering amplitudes are perturbatively calculable.

I will next present numerical values for the
DPE-to-jets cross section.
For comparison purposes, I will also give
(a) the
inclusive two-jet cross section
(i.e., without a diffractive requirement:
$A+B\to\mbox{\rm jet}+\mbox{\rm jet}+X)$, and
(b) the result of applying the
Ingelman-Schlein model to DPE, which gives a result for the
process $A+B\to\mbox{\rm $A'+B'+$jet}+\mbox{\rm jet}+X$. This process we call
factorized double-pomeron-exchange (FDPE).

The total cross
sections integrated over $y_{+}$, $y_{-}$ and for $E_T>5.0~\GeV$
are, at $\sqrt {s}=1800~\GeV$,
\begin{eqnarray}
\sigma_{\rm incl}(1800,5)&=&2.4~\mb,\nonumber\\
\sigma_{\rm FDPE}(1800,5)&=&0.0022~\mb,\nonumber\\
\sigma_{\rm NDPE}(1800,5)&=&0.17~\mb,\nonumber\\
\label{sigma.1800}\end{eqnarray}
and at $\sqrt {s}=630~\GeV$,
\begin{eqnarray}
\sigma_{\rm incl}(630,5)&=&0.31~\mb,\nonumber\\
\sigma_{\rm FDPE}(630,5)&=&0.000062~\mb,\nonumber\\
\sigma_{\rm NDPE}(630,5)&=&0.044~\mb.\nonumber\\
\label{sigma.630}\end{eqnarray}
The integration range for the rapidities $y_{+}$ and $y_{-}$ was restricted
so that both incoming partons into the hard process had momentum fractions
less than 0.05.
This corresponds to
a typical selection cut on the diffractive
hadron for identifying pomeron exchange events.
The same
cut on the incoming partons was made for the estimate of
the inclusive jet cross sections.  If one allows the complete range of
momentum fractions from 0 to 1 for this case one obtains
$\sigma_{{\rm i}{\rm n}{\rm c}{\rm l}}(1800,5)=5.0~\mb$
and $\sigma_{{\rm i}{\rm n}{\rm c}{\rm l}}(630,5)=1.3~\mb$.
For quark jets in the NDPE case, we have verified that their cross sections
are at least a few orders of magnitude smaller.
Also for the NDPE case,
by a simple hand calculation of the total
cross section using Eqs.\ (\ref{dsig2}) and (\ref{forcx})
one can confirm the
order of magnitudes quoted above.  From this
exercise, the factor 4 increase at
$\sqrt{s}=1800 ~\GeV$ compared to $\sqrt{s}= 630 ~\GeV$
is seen to arise primarily from the same factor increase in the
accessible region of jet rapidity.

In conclusion I have presented estimates for jet production in 
DPE processes, by a mechanism in
which the whole of the momentum of the pomerons goes into the jets.
The process has a quite dramatic signature: the final state consists
of the two diffracted hadrons, two high-$E_T$ jets, and {\em
nothing} else.  It is a leading-twist contribution, not suppressed
by a power of $E_T$, despite the fact that the pomeron is a composite
object.  This is permitted because the factorization theorem does not
apply to diffractive processes.

The model presented here is, of course, quite primitive.  However, by
being a complete and consistent calculation in lowest
order perturbation theory, it does establish an important
principle:  that one cannot treat the exchanged pomeron
as a particle with associated parton densities.  The
approximations we have made are concerned only with
extracting the leading power behavior in the appropriate
kinematic limit.
Since gauge-invariance causes
some quite non-trivial cancelations before the final
result is obtained, the exact gauge invariance and
consistency of our model is critical to establishing the
general principles.

If further examination supports the sizes of cross sections we
predict, then the mechanism presented here could be very important for
all kinds of studies.  For example, one can produce the Higgs boson
\cite{BL,Bj}.
Although the cross section would be a lot lower than the total Higgs
cross section, the lack of a background event could make up for the
lower rate in terms of the usefulness of the signature, at least for
certain ranges of parameters.

\medskip

This work was supported in part by the U.S. Department
of Energy under grant numbers DE-FG02-90ER-40577 and
DE-FG02-93ER40771.

Figure 1 Caption:
      Our model of the
      non-factorizing Double-Pomeron-Exchange (NDPE) amplitude
      with two gluon jets produced.

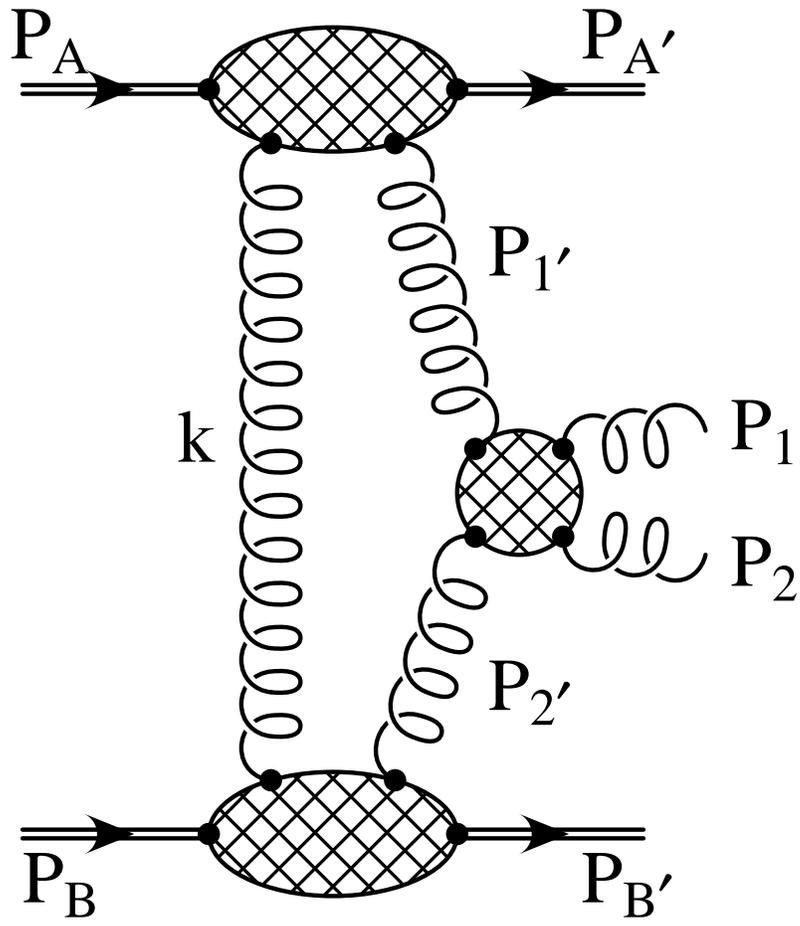
\end{document}